\documentclass[10pt]{article}

\usepackage{geometry}	
\usepackage{amsmath}
\usepackage{amssymb}
\usepackage{amsthm}
\usepackage{mathrsfs}
\usepackage{enumerate}
\usepackage{epsf}
\usepackage{psfrag}
\usepackage{dsfont}
\usepackage{hyperref}
\usepackage{cleveref}
\usepackage{graphicx}
\usepackage{empheq}
\usepackage{hyperref}
\usepackage{epsfig}  
\usepackage{pstool}
\DeclareGraphicsExtensions{.eps,.art,.ART,.ps}
\allowdisplaybreaks
\newcommand{\bbR}{\mathbb{R}}      

\newtheorem{Thm}{Theorem}[section]

\begin{document}
\title{Test fields cannot destroy extremal de Sitter black holes}
\author{Jos\'e Nat\'ario$^1$ and Rodrigo Vicente$^{2}$\\
{\small $^1$ CAMGSD, Departamento de Matem\'{a}tica, Instituto Superior T\'{e}cnico, Universidade de Lisboa, Portugal}\\ 
{\small $^2$ CENTRA, Departamento de F\'{\i}sica, Instituto Superior T\'{e}cnico, Universidade de Lisboa, Portugal}}
\date{}
\maketitle
\begin{abstract}
We determine the timelike Killing vector field that gives the correct definition of energy for test fields propagating in a Kerr-Newman-de Sitter spacetime, and use this result to prove that test fields cannot destroy extremal Kerr-Newman-de Sitter black holes.
\end{abstract}
%
%
%
%
%
\section{Introduction}\label{section0}
In a famous paper \cite{W74}, Wald tested the weak cosmic censorship conjecture \cite{P69, W97} by dropping charged and/or spinning test particles into the event horizon of an extremal Kerr-Newman black hole. Both him and subsequent authors \cite{TdFC76, Needham80} found that the particle would not go in for values of the conserved quantities (energy, angular momentum, charge and/or spin) which would overspin/overcharge the black hole. Similar conclusions were reached by analyzing scalar and electromagnetic test fields propagating in extremal Kerr-Newman black hole backgrounds \cite{Semiz11, Toth12, DS13, Duztas14}. In this case, the fluxes of energy, angular momentum and charge across the event horizon were found to be always insufficient to overspin/overcharge the black hole. Some of these results were extended to higher dimensions \cite{BCNR10, RV17, ASZZ17} and also to the case when there is a negative cosmological constant \cite{GL16, RS14, Gwak18}. At the same time, it was noticed that Wald's thought experiment might produce naked singularities when applied to nearly extremal black holes \cite{Hubeny99, MS07, JS09, SS11}. However, in this case the perturbation could not be assumed to be infinitesimal, and so backreaction effects would have to be taken into account; when this was done, the validity of the cosmic censorship conjecture appeared to be restored \cite{Hod08, BCK10, ZVPH13, SPAJ15, CBSM15}.

In~\cite{NQV16}, the authors (in collaboration with L.\ Queimada) gave a general argument showing that extremal Kerr-Newman and Kerr-Newman-anti-de Sitter black holes cannot be overspun/overcharged by any type of test matter satisfying the null energy condition at the event horizon. This argument was later extended by Sorce and Wald \cite{SW17} to the case of quasi-extremal Kerr-Newman black holes by considering the second order variation of the black hole mass.

In all these gedanken experiments, however, one must be very careful with what is meant by the energy of the test matter, and how it relates to the increase in the black hole mass. In fact, from a logical point of view, these are independent concepts: the energy of the test matter is computed with respect to a given timelike Killing vector field, whereas the black hole mass is a parameter in a black hole solution of the Einstein-Maxwell field equations. In the asymptotic flat case, the two can be related via the ADM mass: indeed, the ADM mass of a spacetime containing an isolated black hole is precisely the black hole mass, whereas the energy of test matter located in the asymptotically flat region (measured with respect to the unique timelike Killing vector field) simply adds to the ADM mass; since this energy is conserved as the test matter moves into the black hole spacetime, the black hole mass should increase by precisely that amount when the test matter is absorbed. In the non-asymptotically flat cases, however, there is no ADM mass, and there may exist many or no timelike Killing vector fields in the asymptotic region. In the asymptotically anti-de Sitter (AdS) case there are notions of total mass available \cite{Wang01, CN02}, and these were used in \cite{NQV16}, together with the results in \cite{Olea05}, to determine which of the infinitely many stationary Killing fields should be used to compute the energy of the test matter.\footnote{This Killing vector field turns out to be the one corresponding to the zero angular momentum observers at infinity.} Notice that this choice is critical,\footnote{The approach by Sorce and Wald \cite{SW17} sidesteps this difficulty by deriving a formula for the mass variation directly in terms of the event horizon Killing generator, but this has only been done in the asymptotically flat case.} and in fact incorrect choices have lead to erroneous claims of violations of weak cosmic censorship in the literature, as pointed out in \cite{MO15}; such claims have been disproved by \cite{GL16}. In the asymptotically de Sitter (dS) case, on the other hand, there exists neither a generally accepted notion of total mass (see however \cite{KT02, LXZ10}) nor a Killing vector field which is timelike in the asymptotic region, and so it is not clear how one should compute the energy of the test matter falling into the black hole. The main purpose of the present paper is to address this issue, and, as a result, to extend the results in \cite{NQV16} to asymptotically dS black holes. As an added bonus, we will confirm that the choice of timelike Killing vector field in \cite{NQV16} for the asymptotically AdS case is indeed correct.

The strategy that we will employ is the following: by letting the mass parameter (Section~\ref{section1}) and also the charge parameter (Section~\ref{section2}) become functions of the radial coordinate $r$, we construct a metric that interpolates between two Kerr-Newman-(A)dS regions of different (physical) masses $M_1$ and $M_2$. The energy-momentum tensor of the (unphysical) field generating this metric can be computed from the Einstein equations, and the corresponding energy can be calculated with respect to any given timelike Killing vector field. It turns out that this energy, possibly corrected by the electromagnetic field energy (Section~\ref{section2}), is precisely the difference $M_2-M_1$ between the two physical masses for a particular choice of Killing vector field (coinciding with the choice in \cite{NQV16}, in the asymptotically AdS case). This result is then used in Section~\ref{section3} to argue that the increase in the black hole physical mass when absorbing test matter is always equal to the matter energy computed with respect to this specific Killing vector field, whose uniqueness is discussed in Section~\ref{section4}. Finally, Section~\ref{section5} contains a proof that extremal Kerr-Newman-dS black holes cannot be destroyed by test matter.

We follow the conventions of \cite{MTW73, W84}; in particular, we use a system of units for which $c=G=1$. We used {\sc Mathematica} for symbolic and numeric computations.
%
%
%
\section{Kerr-(A)dS}\label{section1}
In this section we construct a metric that interpolates between two Kerr-(A)dS regions of different (physical) masses $M_1$ and $M_2$ by letting the mass parameter become a function of the radial coordinate $r$. We then determine, from the Einstein equations, the energy-momentum tensor of the (unphysical) field generating this metric, and use it to compute the corresponding energy with respect to a given timelike Killing vector field. This energy is seen to be precisely the difference $M_2-M_1$ between the two physical masses for a particular choice of Killing vector field.

We start by recalling the Kerr-Newman-(A)dS metric, given in Boyer-Lindquist coordinates by
\begin{align}
ds^2 = & - \frac{\Delta_r}{\rho^2}\left( dt - \frac{a \sin^2 \theta}{\Xi} d\varphi \right)^2 + \frac{\rho^2}{\Delta_r} dr^2 \nonumber \\
& + \frac{\rho^2}{\Delta_\theta} d \theta^2 + \frac{\Delta_\theta \sin^2\theta}{\rho^2}\left( a \, dt - \frac{r^2 + a^2}{\Xi} d\varphi \right)^2, \label{KNAdS}
\end{align}
where 
\begin{align}
& \rho^2 = r^2 + a^2 \cos^2 \theta; \\
& \Xi = 1 \pm \frac{a^2}{l^2}; \\
& \Delta_r = (r^2 + a^2)\left(1\mp \frac{r^2}{l^2}\right) - 2mr + q^2; \\
& \Delta_\theta = 1 \pm \frac{a^2}{l^2}\cos^2\theta
\end{align}
(see for instance \cite{CCK00}). In what follows, the upper sign will always refer to a positive cosmological constant, and the lower sign to a negative cosmological constant, given in terms of the parameter $l$ by\footnote{Note that the Kerr-Newman metric can be obtained by taking the limit $l^2\to\infty$.}
\begin{equation}
\Lambda = \pm \frac{3}{l^2}.
\end{equation}
The mass, spin and electric charge parameters are denoted by $m$, $a$ and $q$, respectively; these parameters are related to the so-called physical mass $M$, angular momentum $J$ and electric charge $Q$ by
\begin{equation}\label{physical}
M = \frac{m}{\Xi^2}, \qquad J = \frac{ma}{\Xi^2}, \qquad Q = \frac{q}{\Xi}.
\end{equation}
Together with the electromagnetic $4$-potential
\begin{equation}\label{empot}
A = - \frac{qr}{\rho^2}\left(dt - \frac{a\sin^2\theta}{\Xi} d\varphi\right),
\end{equation}
the Kerr-Newman-(A)dS metric is a solution of the Einstein-Maxwell equations with cosmological constant $\Lambda$. It admits a two-dimensional group of isometries, generated by the Killing vector fields $X = \frac{\partial}{\partial t}$ and $Y=\frac{\partial}{\partial \varphi}$. 

\begin{figure}[h!]
\begin{center}
\psfrag{r1}{$r_1$}
\psfrag{r2}{$r_2$}
\psfrag{m=m1}{$m=m_1$}
\psfrag{m=m2}{$m=m_2$}
\psfrag{m=m(r)}{$m=m(r)$}
\epsfxsize=0.6\textwidth
\leavevmode
\epsfbox{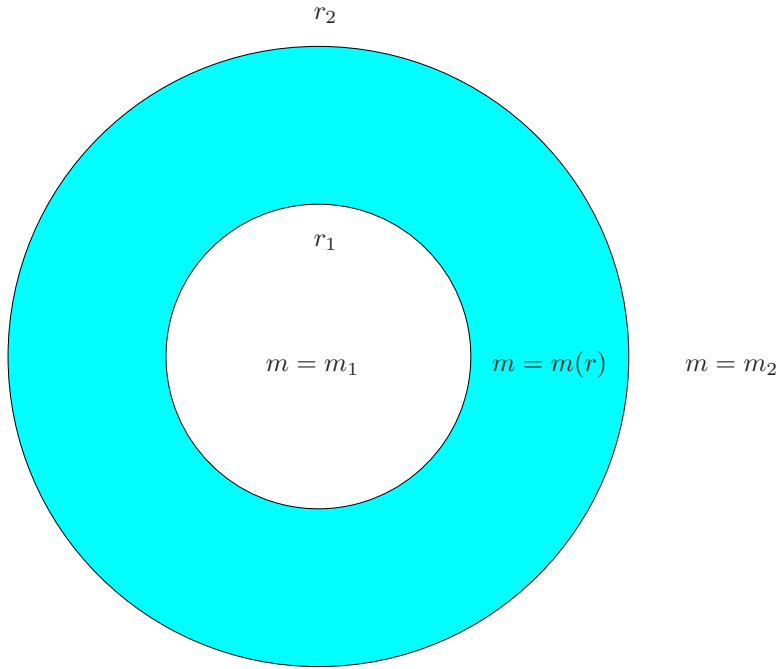}
\end{center}
\caption{Schematic diagram for the spacetime interpolating between Kerr-Newman-(A)dS metrics with different masses.} \label{interpolating}
\end{figure}

In this section we will focus on the simplest case of an electrically neutral Kerr-(A)dS spacetime, corresponding to $q=0$. We consider the stationary spacetime constructed as follows (Figure~\ref{interpolating}): for $r \leq r_1$ it coincides with a Kerr-(A)dS solution with mass parameter $m_1$; for $r \geq r_2 > r_1$ it corresponds to a Kerr-(A)dS solution with mass parameter $m_2 > m_1$; and for $r_1 < r < r_2$ it is the solution of the Einstein equations obtained by taking $m=m(r)$ (and $q=0$) in \eqref{KNAdS}, corresponding to some (unphysical) field which generates the energy-momentum tensor $T^{\mu\nu}$ dictated by the Einstein equations. We assume that $r_1$ is larger than the radius of the event horizon corresponding to the mass parameter $m_1$, and that $r_2$ is smaller than the radius of the cosmological horizon corresponding to the mass parameter $m_2$ in the Kerr-dS case. In other words, we take the metric \eqref{KNAdS} with $m=m(r)$ satisfying $m(r)\equiv m_1$ for $r\leq r_1$, $m(r)\equiv m_2$ for $r\geq r_2$, and $\Delta_r(r)>0$ for $r_1 \leq r \leq r_2$; to avoid thin shells, we assume that $m(r)$ is at least $C^1$, implying in particular that $m'(r_1)=m'(r_2)=0$. For this spacetime it is fairly obvious what the energy of the field should be: since the physical masses, $M_1=\dfrac{m_ 1}{\Xi^2}$ and $M_2=\dfrac{m_2}{\Xi^2}$, correspond to the total energy contained in the regions $r<r_1$ and $r<r_2$, respectively, the energy of the field should be $E=\Delta M\equiv M_2-M_1$. We would like to calculate this energy as an integral on a given spacelike hypersurface $S$ extending from $r=r_1$ to $r=r_2$. In fact, it turns out that this is possible in Kerr-AdS, where it is known that (at least for test fields)
\begin{equation}\label{energyKerr}
E=\int_S T^{\mu \nu}K_\mu N_\nu dV_3\,,
\end{equation}
with $N$ the future-pointing unit normal to $S$, and $K$ the Killing vector field
\begin{equation}\label{killing}
K=X-\frac{a}{l^2}Y\,.
\end{equation}
It is interesting to note that $K$ has zero rotation with respect to the zero-angular momentum observers at infinity. There are some works in the literature (e.g.~\cite{Gwak18, Gwak18b}) where an expression analogous to Eq.~\eqref{energyKerr} is used to calculate the energy of test fields propagating on Kerr-dS, but, this time, using the Killing vector field 
\begin{equation}\label{killing2}
K=X+\dfrac{a}{l^2}Y. 
\end{equation}
However, to the best of our knowledge, there is neither a rigorous proof nor a clear physical motivation for the use of this definition of energy. In what follows we will show that, in our particular setup, the definition of Eq.~\eqref{energyKerr} gives $\Delta M$ in both asymptotically AdS and dS spacetimes, if one uses the corresponding Killing vector field $K$, defined by either \eqref{killing} or \eqref{killing2}.

Since the metric $g_{\mu\nu}$ is known, the energy-momentum tensor $T^{\mu\nu}$ of the field is obtained from the Einstein equations as
\begin{equation} \label{stresstensor}
	T^{\mu\nu}=\frac{1}{8 \pi}(G^{\mu\nu} + \Lambda g^{\mu\nu})\,,
\end{equation}
where $G^{\mu\nu}$ is the Einstein tensor. Computing $G^{\mu\nu}$ explicitly, and substituting the last expression in Eq.~\eqref{energyKerr}, we obtain
\begin{equation}\label{energyr}
	E=\int_{r_1}^{r_2} \left[\mathcal{A}(r) m'(r)+ \mathcal{B}(r) m''(r)\right]dr\,,
\end{equation}
where we have chosen a hypersurface $S$ of constant $t$ extending from $r_1$ to $r_2$, and performed the integrations in $\theta$ and $\varphi$. The radial functions $\mathcal{A}$ and $\mathcal{B}$ are given by
 \begin{align}
 	\mathcal{A}(r)&= \mp \frac{l^2}{ a (a^2\pm l^2)^2}\left[2 a (r^2\mp l^2) -   \arctan\left(\frac{a}{r}\right)r\left( a^2 \mp l^2+2r^2\right)\right]\,,\\
 	\mathcal{B}(r)&= \mp \frac{l^2}{2 a (a^2\pm l^2)^2}\left[a r \left(r^2 \mp l^2\right)-\arctan\left(\frac{a}{r}\right)\left(a^2+r^2\right)\left(r^2 \mp l^2\right) \right]\,.
 \end{align}
Integrating Eq.~\eqref{energyr} by parts, we obtain
\begin{align}
	E&=\int_{r_1}^{r_2} \left[\mathcal{B}''(r)-\mathcal{A}'(r)\right] m(r) dr +\biggl[\left(\mathcal{A}(r)-\mathcal{B}'(r)\right)m(r)+\mathcal{B}(r) m'(r)\biggr]_{r_1}^{r_2}\,.
\end{align}
Using $\mathcal{B}''(r)=\mathcal{A}'(r)$, $m'(r_2)=m'(r_1)=0$, and $\mathcal{A}(r)-\mathcal{B}'(r)=\dfrac{1}{\Xi^2}$, the last expression becomes
\begin{equation} \label{energyKerrconsistent}
	E=\frac{m_2-m_1}{\Xi^2}=M_2-M_1\equiv \Delta M \, ,
\end{equation}
as we wanted to show. 

We can also calculate the field angular momentum $L$ as an integral on a given spacelike hypersurface $S$ extending from $r=r_1$ to $r=r_2$. This can be done in Kerr-AdS (at least for test fields), where it is known that
\begin{equation}\label{angularmKerr}
	L=-\int_S T^{\mu \nu}Y_\mu N_\nu dV_3
\end{equation}
(note the minus sign in the integral, since we are using the future-pointing unit timelike normal but now the Killing vector field is spacelike). In our particular setup, we know what the angular momentum of the field should be: since the physical angular momenta, $J_1=a M_1$ and $J_2=a M_2$, correspond to the total angular momentum contained in the regions $r<r_1$ and $r<r_2$, respectively, the angular momentum of the field should be $L=\Delta J\equiv J_2-J_1$. We will now show that, in our setup, the definition of Eq.~\eqref{angularmKerr} does indeed give $\Delta J$ in both asymptotically AdS and dS spacetimes.
Computing $G^{\mu\nu}$ explicitly, and substituting Eq.~\eqref{stresstensor} in the definition of Eq.~\eqref{angularmKerr}, we obtain
\begin{equation}\label{angulamr}
	L=\int_{r_1}^{r_2}\left[\mathcal{C}(r) m'(r)+ \mathcal{D}(r) m''(r)\right]dr\,,
\end{equation}
where again we have chosen a hypersurface $S$ of constant $t$ extending from $r_1$ to $r_2$, and performed the integrations in $\theta$ and $\varphi$. The radial functions $\mathcal{C}$ and $\mathcal{D}$ are given by 
 \begin{align}
\mathcal{C}(r)&= 2 l^4 \frac{a^2+r^2}{ a^2 (a^2\pm l^2)^2}\left[a-r  \arctan\left(\frac{a}{r}\right)\right]\,,\\
\mathcal{D}(r)&= l^4 \frac{a^2+r^2}{2 a^2 (a^2\pm l^2)^2}\left[a r -\arctan\left(\frac{a}{r}\right)\left(a^2+r^2\right) \right]\,.
\end{align}
Integrating Eq.~\eqref{angulamr} by parts, we obtain
\begin{align}
L&=\int_{r_1}^{r_2}\left[\mathcal{D}''(r)-\mathcal{C}'(r)\right]m(r)dr+\biggl[\left(\mathcal{C}(r)-\mathcal{D}'(r)\right)m(r)+\mathcal{D}(r) m'(r)\biggr]_{r_1}^{r_2}\,.
\end{align}
Using $\mathcal{D}''(r)=\mathcal{C}'(r)$, $m'(r_2)=m'(r_1)=0$, and $\mathcal{C}(r)-\mathcal{D}'(r)=\dfrac{a}{\Xi^2}$, the last expression becomes
\begin{equation} \label{angularmKerrconsistent}
L=a\frac{m_2-m_1}{\Xi^2}=a (M_2-M_1)\equiv \Delta J \, ,
\end{equation}
as we wanted to show.  As a consequence, the energy of the unphysical field computed by using any timelike Killing vector field of the form 
\begin{equation}
K + \omega Y = X + \left(\omega \pm \frac{a}{l^2}\right) Y
\end{equation}
is
\begin{equation}
E + \omega L = (1+\omega a) \Delta M,
\end{equation}
strongly suggesting that $K$ (that is, $\omega=0$) is in fact the correct choice. We will have more to say about the uniqueness of $K$ in Section~\ref{section4}.
%
%
%
\section{Kerr-Newman-(A)dS}\label{section2}
In this section we construct a metric that interpolates between two Kerr-Newman-(A)dS regions of different (physical) masses $M_1$ and $M_2$ and (physical) charges $Q_1$ and $Q_2$ by letting both the mass and the charge parameters become functions of the radial coordinate $r$. We then determine, from the Einstein equations, the energy-momentum tensor of the (unphysical) field generating this metric, and use it to compute the corresponding energy with respect to a given timelike Killing vector field. This energy, appropriately corrected by the electromagnetic field energy, is seen to be precisely the difference $M_2-M_1$ between the two physical masses for the particular choice of Killing vector field given by Eqs.~\eqref{killing} and \eqref{killing2}, thus generalizing the results in Section~\ref{section1}.

\begin{figure}[h!]
\begin{center}
\psfrag{i+}{$i^+$}
\psfrag{i-}{$i^-$}
\psfrag{H+}{$\mathscr{H^+}$}
\psfrag{H-}{$\mathscr{H^-}$}
\psfrag{CH+}{$\mathscr{CH^+}$}
\psfrag{CH-}{$\mathscr{CH^-}$}
\psfrag{I+}{$\mathscr{I^+}$}
\psfrag{I-}{$\mathscr{I^-}$}
\psfrag{S}{$\Sigma$}
\epsfxsize=0.6\textwidth
\leavevmode
\epsfbox{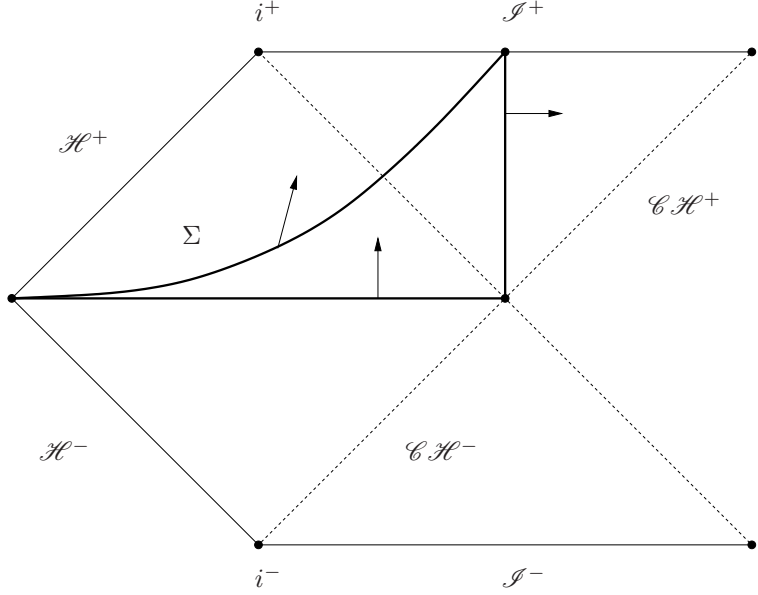}
\end{center}
\caption{Penrose diagram illustrating the deformation of an unbounded spacelike hypersurface $\Sigma$ into the union of two hypersurfaces of constant $t$, with the corresponding unit normals depicted.} \label{Penrose}
\end{figure}

Let us then take the charge parameter $q(r)$ to be changing in the region $r_1<r<r_2$, with $q(r)\equiv q_1$ for $r \leq r_1$ and $q(r)\equiv q_2$ for $r \geq r_2$. Moreover, assume that $q'(r_1)= q'(r_2)=0$, and again that $\Delta_r(r)>0$ for $r_1\leq r\leq r_2$. In this case we have an electromagnetic field with energy-momentum tensor $T^{\mu\nu}_\text{EM}$, and it is not obvious what the mass contained on a spacelike hypersurface $S$ extending from $r_1$ to $r_2$ should be. In the asymptotically flat case, it is well known that the physical mass accounts also for the electromagnetic energy in the whole spacetime. By analogy, the mass contained on a spacelike hypersurface $S$ extending from $r_1$ to $r_2$ should then be
\begin{equation} \label{EKNAsS}
E=\left(M_2-\int_{r>r_2} T_\text{EM,2}^{\mu \nu}K_\nu N_\mu dV_3\right)-\left(M_1-\int_{r>r_1} T_\text{EM,1}^{\mu \nu}K_\nu N_\mu dV_3\right)\,,
\end{equation}
where the first term is the mass contained in $r<r_2$, and the second term is the mass in $r<r_1$. Here, $T^{\mu\nu}_\text{EM,1}$ and $T^{\mu\nu}_\text{EM,2}$ are the energy-momentum tensors of the electromagnetic field in a Kerr-Newman-(A)dS spacetime with mass parameters $m_1$ and $m_2$, and charge parameters $q_1$ and $q_2$, respectively. Note that in \eqref{EKNAsS} we have already made use of the Killing vector field $K$ to calculate the electromagnetic energy. On the other hand, the energy contained on $S$ should be directly
\begin{equation} \label{energyKerrN}
E=\int_S \left(T^{\mu \nu}+T_\text{EM}^{\mu \nu}\right)K_\mu N_\nu dV_3\,,
\end{equation}
where $T^{\mu\nu}_\text{EM}$ is the energy-momentum tensor of the electromagnetic field in the Kerr-Newman-(A)dS spacetime with varying mass parameter $m(r)$ and varying charge parameter $q(r)$. Thus, if our definition of energy is to be consistent, we must have
\begin{align}\label{deltam}
	\Delta M=\int_S \left(T^{\mu \nu}+T_\text{EM}^{\mu \nu}\right)K_\mu N_\nu dV_3+\int_{r>r_2} T_\text{EM,2}^{\mu \nu}K_\nu N_\mu dV_3-\int_{r>r_1} T_\text{EM,1}^{\mu \nu}K_\nu N_\mu dV_3 \,.
\end{align}
Again, since the metric $g_{\mu\nu}$ is known, the Einstein equations imply that
\begin{equation} \label{stresstensorem}
	T^{\mu\nu}+T^{\mu\nu}_\text{EM}=\frac{1}{8 \pi}\left(G^{\mu\nu}+\Lambda g^{\mu\nu}\right)\,.
\end{equation}
Computing $G^{\mu\nu}$ explicitly, and using Eq.~\eqref{stresstensorem}, allows us to write the first integral in Eq.~\eqref{deltam} as
\begin{align} \label{FirstTermEn}
	\int_S \left(T^{\mu \nu}+T_\text{EM}^{\mu \nu}\right)&K_\mu N_\nu dV_3=\nonumber\\ &\int_{r_1}^{r_2} dr \left[\mathcal{A}(r) m'+ \mathcal{B}(r) m''+\mathcal{E}(r) q^2-r \mathcal{E}(r) (q^2)'-\frac{\mathcal{B}(r)}{2 r}(q^2)''\right]\,,
\end{align}
where again we have chosen a hypersurface $S$ of constant $t$ extending from $r_1$ to $r_2$, and performed the integrations in $\theta$ and $\varphi$. The radial functions $\mathcal{A}$ and $\mathcal{B}$ are defined as in the last section, and
\begin{equation}
	\mathcal{E}(r)= \mp \frac{l^2}{2 a r^3 (a^2 \pm l^2)^2}\left[a r \left(r^2 \mp l^2\right)-\arctan\left(\frac{a}{r}\right)\left(r^4\pm a^2 l^2\right) \right]\, .
\end{equation}
Integrating by parts, and using the results of the last section, we have
\begin{align} 
	\int_S \left(T^{\mu \nu}+T_\text{EM}^{\mu \nu}\right)&K_\mu N_\nu dV_3 = \nonumber\\&\Delta M+\int_{r_1}^{r_2} dr \left[\mathcal{E}(r)+\left(r \mathcal{E}(r)\right)'-\left(\frac{\mathcal{B}(r)}{2 r}\right)'' \right]q^2+  \left[\left(\left[\frac{\mathcal{B}(r)}{2 r}\right]'-r \mathcal{E}(r)\right)q^2-\frac{\mathcal{B}(r)}{2 r}(q^2)'\right]_{r_1}^{r_2} \,.
\end{align}
Using $q'(r_1)= q'(r_2)=0$, and
\begin{equation}\label{primc}
	\mathcal{E}(r)=\left[\left(\frac{\mathcal{B}(r)}{2 r}\right)'-r \mathcal{E}(r)\right]'\,,
\end{equation}
we obtain
\begin{equation}
	\int_S \left(T^{\mu \nu}+T_\text{EM}^{\mu \nu}\right)K_\mu N_\nu dV_3=\Delta M+\left[\left(\left[\frac{\mathcal{B}(r)}{2 r}\right]'-r \mathcal{E}(r)\right)q^2\right]_{r_1}^{r_2} \,.
\end{equation}
Furthermore, the last two terms of Eq.~\eqref{deltam} are
\begin{align} \label{last}
\int_{r>r_2} T_\text{EM,2}^{\mu \nu}K_\nu N_\mu dV_3&-\int_{r>r_1} T_\text{EM,1}^{\mu \nu}K_\nu N_\mu dV_3=(q_2)^2\int_{r_2}^{\infty} dr \,\mathcal{E}(r)-(q_1)^2\int_{r_1}^{\infty} dr \,\mathcal{E}(r)\,,
\end{align}
where we have used Eq.~\eqref{FirstTermEn}, with $m\equiv m_ 2$ ($m\equiv m_1$), $q\equiv q_2$ ($q\equiv q_1$) in the first (second) term, but integrating on a spacelike hypersurface of constant $t$ with $r>r_2$ ($r>r_1$).  In the Kerr-Newman-dS case, a hypersurface of constant $t$ is not spacelike beyond the cosmological horizon; nevertheless, since we are integrating a divergenceless quantity, any unbounded spacelike hypersurface can be deformed into the union of a spacelike hypersurface of constant $t$ within the cosmological horizon and a timelike hypersurface of constant $t$ beyond the cosmological horizon (see Figure~\ref{Penrose}).

Using Eq.~\eqref{primc}, Eq.~\eqref{last} becomes
\begin{align}
	&\int_{r>r_2} T_\text{EM,2}^{\mu \nu}K_\nu N_\mu dV_3-\int_{r>r_1} T_\text{EM,1}^{\mu \nu}K_\nu N_\mu dV_3=\nonumber\\&(q_2)^2\left[\left(\left[\frac{\mathcal{B}(r)}{2 r}\right]'-r \mathcal{E}(r)\right)\right]_{r_2}^{\infty}-(q_1)^2\left[\left(\left[\frac{\mathcal{B}(r)}{2 r}\right]'-r \mathcal{E}(r)\right)\right]_{r_1}^{\infty}=-\left[\left(\left[\frac{\mathcal{B}(r)}{2 r}\right]'-r \mathcal{E}(r)\right)q^2\right]_{r_1}^{r_2}\,,
\end{align}
where in the last equality we used 
\begin{equation*}
	\lim_{r\to \infty}\left(\left[\frac{\mathcal{B}(r)}{2 r}\right]'-r \mathcal{E}(r)\right)=0\,.
\end{equation*}
Putting everything together, we finally obtain
\begin{equation} \label{energyKerrNconsistent}
	\int_S \left(T^{\mu \nu}+T_\text{EM}^{\mu \nu}\right)K_\mu N_\nu dV_3+\int_{r>r_2} T_\text{EM,2}^{\mu \nu}K_\nu N_\mu dV_3-\int_{r>r_1} T_\text{EM,1}^{\mu \nu}K_\nu N_\mu dV_3=\Delta M\,,
\end{equation}
showing that our definition of energy is indeed consistent.

In the same way, the angular momentum contained on $S$ should be
\begin{equation}
L=\left(J_2+\int_{r>r_2} T_\text{EM,2}^{\mu \nu}Y_\nu N_\mu dV_3\right)-\left(J_1+\int_{r>r_1} T_\text{EM,1}^{\mu \nu}Y_\nu N_\mu dV_3\right)\,,
\end{equation}
where the first term is the angular momentum contained in $r<r_2$, and the second term is the angular momentum contained in $r<r_1$ (note the minus sign in the integral, since we are using the future-pointing unit timelike normal but now the Killing vector field is spacelike). On the other hand, the angular momentum contained on $S$ should be directly
\begin{equation} \label{angularmKerrN}
L=-\int_S \left(T^{\mu \nu}+T_\text{EM}^{\mu \nu}\right)Y_\mu N_\nu dV_3\,.
\end{equation}
Thus, if this definition of angular momentum is to be consistent, the relation
\begin{align}\label{deltaj}
\Delta J=-\int_S \left(T^{\mu \nu}+T_\text{EM}^{\mu \nu}\right)Y_\mu N_\nu dV_3-\int_{r>r_2} T_\text{EM,2}^{\mu \nu}Y_\nu N_\mu dV_3+\int_{r>r_1} T_\text{EM,1}^{\mu \nu}Y_\nu N_\mu dV_3
\end{align}
must hold. Computing $G^{\mu\nu}$ explicitly, and using Eq.~\eqref{stresstensorem}, allows us to write the first integral in Eq.~\eqref{deltaj} as
\begin{align} \label{FirstTermAn}
-\int_S \left(T^{\mu \nu}+T_\text{EM}^{\mu \nu}\right)&Y_\mu N_\nu dV_3=\nonumber\\ &\int_{r_1}^{r_2} dr \left[\mathcal{C}(r) m'+ \mathcal{D}(r) m''+\mathcal{F}(r) q^2-r \mathcal{F}(r) (q^2)'-\frac{\mathcal{D}(r)}{2 r}(q^2)''\right]\,,
\end{align}
where again we have chosen a hypersurface $S$ of constant $t$ extending from $r_1$ to $r_2$, and performed the integrations in $\theta$ and $\varphi$. The radial functions $\mathcal{C}$ and $\mathcal{D}$ are defined as in the last section, and
\begin{equation}
\mathcal{F}(r)= l^4 \frac{a^2+r^2}{2 a^2 r^3 (a^2 \pm l^2)^2}\left[a r +\arctan\left(\frac{a}{r}\right)\left(a^2- r^2\right) \right]\,.
\end{equation}
Integrating by parts, and using the results in the last section, we have
\begin{align} 
-\int_S &\left(T^{\mu \nu}+T_\text{EM}^{\mu \nu}\right)Y_\mu N_\nu dV_3=\nonumber\\&\Delta J+\int_{r_1}^{r_2} dr \left[\mathcal{F}(r)+\left(r \mathcal{F}(r)\right)'-\left(\frac{\mathcal{D}(r)}{2 r}\right)'' \right]q^2+  \left[\left(\left[\frac{\mathcal{D}(r)}{2 r}\right]'-r \mathcal{F}(r)\right)q^2-\frac{\mathcal{D}(r)}{2 r}(q^2)'\right]_{r_1}^{r_2} \,.
\end{align}
Using $q'(r_1)= q'(r_2)=0$, and
\begin{equation}\label{primf}
\mathcal{F}(r)=\left[\left(\frac{\mathcal{D}(r)}{2 r}\right)'-r \mathcal{F}(r)\right]'\,,
\end{equation}
we have
\begin{equation}
-\int_S \left(T^{\mu \nu}+T_\text{EM}^{\mu \nu}\right)Y_\mu N_\nu dV_3=\Delta J+\left[\left(\left[\frac{\mathcal{D}(r)}{2 r}\right]'-r \mathcal{F}(r)\right)q^2\right]_{r_1}^{r_2} \,.
\end{equation}
Moreover, the last two integrals of Eq.~\eqref{deltaj} are
\begin{align}
-\int_{r>r_2} T_\text{EM,2}^{\mu \nu}Y_\nu N_\mu dV_3&+\int_{r>r_1} T_\text{EM,1}^{\mu \nu}Y_\nu N_\mu dV_3=(q_2)^2\int_{r_2}^{\infty} dr \,\mathcal{F}(r)-(q_1)^2\int_{r_1}^{\infty} dr \,\mathcal{F}(r)\,,
\end{align}
where we have used Eq.~\eqref{FirstTermAn}, with $m\equiv m_ 2$ ($m\equiv m_1$), $q\equiv q_2$ ($q\equiv q_1$) in the first (second) term, but integrating on a spacelike hypersurface of constant $t$ with $r>r_2$ ($r>r_1$). 
Using Eq.~\eqref{primf}, the last expression becomes
\begin{align}
&-\int_{r>r_2} T_\text{EM,2}^{\mu \nu}Y_\nu N_\mu dV_3+\int_{r>r_1} T_\text{EM,1}^{\mu \nu}Y_\nu N_\mu dV_3=\nonumber\\&(q_2)^2\left[\left(\left[\frac{\mathcal{D}(r)}{2 r}\right]'-r \mathcal{F}(r)\right)\right]_{r_2}^{\infty}-(q_1)^2\left[\left(\left[\frac{\mathcal{D}(r)}{2 r}\right]'-r \mathcal{F}(r)\right)\right]_{r_1}^{\infty}=-\left[\left(\left[\frac{\mathcal{D}(r)}{2 r}\right]'-r \mathcal{F}(r)\right)q^2\right]_{r_1}^{r_2}\,,
\end{align}
where, in the last equality, we used 
\begin{equation*}
\lim_{r\to \infty}\left(\left[\frac{\mathcal{D}(r)}{2 r}\right]'-r \mathcal{F}(r)\right)=0\,.
\end{equation*}
Putting everything together, we finally obtain
\begin{equation} \label{angularmKerrNconsistent}
-\int_S \left(T^{\mu \nu}+T_\text{EM}^{\mu \nu}\right)Y_\mu N_\nu dV_3-\int_{r>r_2} T_\text{EM,2}^{\mu \nu}Y_\nu N_\mu dV_3+\int_{r>r_1} T_\text{EM,1}^{\mu \nu}Y_\nu N_\mu dV_3=\Delta J\,,
\end{equation}
showing that our definition of angular momentum is indeed consistent. As a consequence, a timelike Killing vector field of the form 
\begin{equation}
K + \omega Y = X + \left(\omega \pm \frac{a}{l^2}\right) Y
\end{equation}
will again only satisfy Eq.~\eqref{energyKerrNconsistent} if $\omega a =0$, strongly suggesting that $K$ (that is, $\omega=0$) is in fact the correct choice. The uniqueness of $K$ will be further discussed in Section~\ref{section4}.
%
%
\section{Linearized calculation}\label{section3}
In the previous sections we showed that there exists a timelike Killing vector field $K$, given by Eqs.~\eqref{killing} and \eqref{killing2}, such that the definitions in Eqs.~\eqref{energyKerr} and \eqref{energyKerrN} give the correct total energy $E$ contained in the (unphysical) field that is generated by allowing the mass and charge parameters to become functions of the radial coordinate. This energy is related to the variation $\Delta M = M_2 - M_1$ of the physical mass by Eqs.~\eqref{energyKerrconsistent} and \eqref{energyKerrNconsistent}. However, the Killing vector field $K$ is defined on a unphysical stationary spacetime that coincides with Kerr-Newman-(A)dS spacetimes of mass and charge parameters $m_1$ and $q_1$ for $r \leq r_1$, and mass and charge parameters $m_2$ and $q_2$ for $r \geq r_2$, whereas our aim is to identify the timelike Killing vector field that gives the correct definition of energy of {\em test fields} on a {\em fixed} Kerr-Newman-(A)dS background. 

\begin{figure}[h!]
\begin{center}
\psfrag{r=r1}{$r=r_1$}
\psfrag{r=r2}{$r=r_2$}
\psfrag{t=t0}{$t=t_0$}
\psfrag{t=t0+1}{$t=t_0 + 1$}
\epsfxsize=0.6\textwidth
\leavevmode
\epsfbox{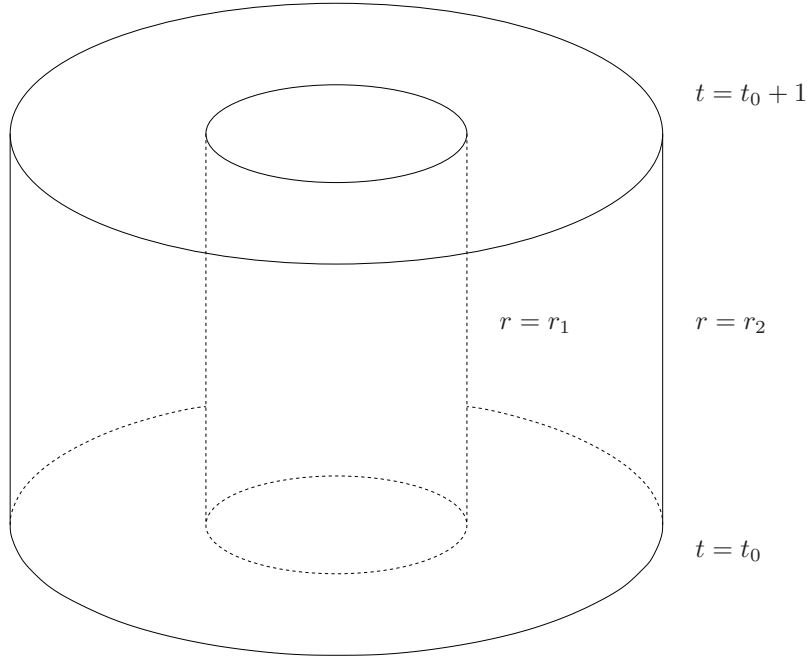}
\end{center}
\caption{Domain for the application of the divergence theorem.} \label{divergence}
\end{figure}

To achieve this goal, we consider a solution of the linearized Einstein-Maxwell equations, possibly coupled to matter, on a Kerr-Newman-(A)dS background of mass and charge parameters $m_1$ and $q_1$, vanishing for $r \leq r_1$ and coinciding with the linearized Kerr-Newman-(A)dS solution of mass and charge parameters $m_2 = m_1 + \Delta m$ and $q_2 = q_1 + \Delta q$ for $r \geq r_2$ (and the same spin parameter $a$); if the energy computed from the linearized energy-momentum tensor with respect to the Killing vector field $K$ (which is now defined on the fixed Kerr-Newman-(A)dS background of mass and charge parameters $m_1$ and $q_1$) is $\Delta M=\Delta m / \Xi^2$ then $K$ does indeed give the correct definition of energy. Note that one such linearized solution, albeit for unphysical matter, can be obtained by linearizing the spacetime constructed in the previous sections; as we have shown, the Killing vector field $K$ does give the correct energy in this case.  A simple application of the divergence theorem then shows that $K$ will give the same energy for any other linearized solution, including solutions corresponding to physical matter fields. Indeed, if $\delta g_{\mu\nu}(t,r,\theta,\varphi)$ is an arbitrary linearized metric, $\delta g_{\mu\nu}^0(r,\theta)$ is the linearization of the metric constructed in the previous sections, and $\rho(t)$ is a smooth function satisfying $\rho(t) \equiv 1$ for $t \leq 0$ and $\rho(t) \equiv 0$ for $t \geq 1$, consider the linearized metric $\rho(t-t_0) \delta g_{\mu\nu} + (1-\rho(t-t_0)) \delta g_{\mu\nu}^0$. The linearized energy-momentum tensor corresponding to this metric has zero divergence in the Kerr-Newman-(A)dS background, coincides with the energy-momentum tensor of the arbitrary linearized metric for $t=t_0$, and with the energy-momentum tensor of $\delta g_{\mu\nu}^0$ for $t=t_0+1$. Moreover, it vanishes for $r \leq r_1$ and it is time-independent for $r \geq r_2$ (so in particular does not depend on the choice of $\delta g_{\mu\nu}$ in those regions). Applying the divergence theorem to the vector field $J_\mu=\left(T^{\mu \nu}+T_\text{EM}^{\mu \nu}\right)K_\nu$ in the hollow cylinder defined by $r_1 \leq r \leq r_2$ and $t_0 \leq t \leq t_0 + 1$ (see Figure~\ref{divergence}), we obtain
\begin{align}
	& \int_{r_1 < r < r_2} \left(T^{\mu \nu}+T_\text{EM}^{\mu \nu}\right)K_\mu N_\nu dV_3 - \int_{r_1 < r < r_2} \left(T_0^{\mu \nu}+T_\text{EM,0}^{\mu \nu}\right)K_\mu N_\nu dV_3 \nonumber \\
	& - \int_{r = r_1} \left(T^{\mu \nu}+T_\text{EM}^{\mu \nu}\right)K_\mu N_\nu dV_3 + \int_{r = r_2} \left(T^{\mu \nu}+T_\text{EM}^{\mu \nu}\right)K_\mu N_\nu dV_3 = 0\,,
\end{align}
where the unit normal $N$ is future-pointing when timelike and outward-pointing when spacelike, and the energy-momentum tensor $T_0^{\mu \nu}+T_\text{EM,0}^{\mu \nu}$ refers to $\delta g_{\mu\nu}^0$. Since the last two integrals do not depend on the choice of $\delta g_{\mu\nu}$, and their sum clearly vanishes when one chooses $\delta g_{\mu\nu} = \delta g_{\mu\nu}^0$ (because the first two integrals cancel in that case), it always vanishes; therefore we obtain
\begin{equation}
	\int_{r_1 < r < r_2} \left(T^{\mu \nu}+T_\text{EM}^{\mu \nu}\right)K_\mu N_\nu dV_3 = \int_{r_1 < r < r_2} \left(T_0^{\mu \nu}+T_\text{EM,0}^{\mu \nu}\right)K_\mu N_\nu dV_3\,,
\end{equation}
showing that $K$ does indeed yield the correct energy for any linearized solution.

%
%
%
\section{Uniqueness of $K$}\label{section4}
We have now identified a timelike Killing vector field $K$ in the Kerr-Newman-(A)dS spacetime, given by Eqs.~\eqref{killing} and \eqref{killing2}, such that the definitions in Eqs.~\eqref{energyKerr} and \eqref{energyKerrN} give the correct total energy $E$ contained in linearized (test) fields. This energy is related to the variation $\Delta M = M_2 - M_1$ of the physical mass by Eqs.~\eqref{energyKerrconsistent} and \eqref{energyKerrNconsistent}. Similarly, the definitions in Eqs.~\eqref{angularmKerr} and \eqref{angularmKerrN} give the correct total angular momentum $L$ in the test fields, which is related to the variation $\Delta J = a \Delta M$ of the angular momentum by Eqs.~\eqref{angularmKerrconsistent} and \eqref{angularmKerrNconsistent}. However, because the variations of energy and angular momentum are related through the spin parameter $a$, which we did not vary, the possibility that $K$ is not unique remains.

To understand this, we note that any other future-pointing timelike Killing vector field can be written in the form
\begin{equation}
\tilde{K} = \gamma \left(K+\epsilon Y\right) \,,
\end{equation}
with $\gamma > 0$ and $\epsilon \in \bbR$ appropriately chosen. Combining Eqs.~\eqref{energyKerrNconsistent} and \eqref{angularmKerrNconsistent}, we see that $\tilde{K}$ will also give the correct total energy $E$ contained in the unphysical field if and only if
\begin{equation}
\gamma \Delta M - \gamma \epsilon \Delta J = \Delta M \Leftrightarrow \gamma (1 - \epsilon a) = 1 \, ,
\end{equation}
that is, if and only if
\begin{equation}
\tilde{K} = \frac{1}{1-a \epsilon(a)}\left[K+\epsilon(a) Y\right] \,,
\end{equation}
where we made it explicit that $\epsilon$ is an unknown function of $a$. To show that $\epsilon(a)$ must be identically zero, and therefore that $K$ is unique, we allow the spin parameter $a$ to become a function of $r$ in the region $r_1\leq r\leq r_2$, while keeping the mass and charge parameters fixed. To perform the linearization, we assume that $a(r)= a_0+\delta a(r)$ varies infinitesimally between $a(r_1)= a_0$ and $a(r_2)= a_0+\Delta a$ (that is, $|\delta a(r)| \ll a_0$). Since in this case the calculations are much more involved than in the previous sections, we assume that all quantities are analytic functions of $a$ and expand them as power series of $a_0$ (and to linear order in $\delta a(r)$). In particular, we have
\begin{equation}
	\epsilon(a)=\sum_{n=0}^{+\infty} \epsilon_n a^n \,.
\end{equation} 
In what follows, we will show that $\epsilon_0=\epsilon_1=0$. Due to the complexity of the calculations, we have not computed the higher order coefficients $\epsilon_n$ with $n\geq 2$, but we expect them to also vanish.

To further simplify calculations we consider only the Kerr-(A)dS case $q_1=q_2=0$. Using the definition of Eq.~\eqref{energyKerr} with the Killing vector field $\tilde{K}(a_0)$, and applying the same procedure of the previous sections, we obtain the radial integral
\begin{equation} \label{energyreps}
E = \int_S T^{\mu \nu}(\tilde{K}(a_0))_\mu N_\nu dV_3=\int_{r_1}^{r_2}dr[\mathcal{G}(r)\delta a'(r)+\mathcal{H}(r)\delta a''(r)]\,,
\end{equation}
with the radial functions
 \begin{align}
\mathcal{G}(r)&= -\frac{1}{15 l^4 r^3}\left[\left(10 \epsilon_0 l^4 m r^3 \mp 10 \epsilon_0 l^2 r^6\right)\nonumber\right.\\ &\left.+a_0 \left(10 \epsilon_0^2 l^4 m r^3\mp 10 \epsilon_0^2 l^2 r^6+10 \epsilon_1 l^4 m r^3\mp 10 \epsilon_1 l^2 r^6+20 l^4 m^2+10 l^4 m r\pm 75 l^2 m r^3\mp 40 l^2 r^4+40 r^6\right)\nonumber\right.\\ &\left.+a_0^2 \left(20 \epsilon_0 \epsilon_1 l^4 m r^3\mp 20 \epsilon_0 \epsilon_1 l^2 r^6+4 \epsilon_0 l^4 m^2+2 \epsilon_0 l^4 m r\mp 5 \epsilon_0 l^2 m r^3\mp 50 \epsilon_0 l^2 r^4+52 \epsilon_0 r^6\right)+ \mathcal{O}(a_0^3)  \right]\,,\\
\mathcal{H}(r)&= \frac{1}{30 l^4 r^2}\left[\left(10 \epsilon_0 l^4 m r^3\pm 5 \epsilon_0 l^2 r^6\right)\nonumber \right.\\&\left.+a_0 \left(10 \epsilon_0^2 l^4 m r^3\pm 5 \epsilon_0^2 l^2 r^6+10 \epsilon_1 l^4 m r^3\pm 5 \epsilon_1 l^2 r^6+20 l^4 m^2+20 l^4 m r-20 l^4 r^2\mp 30 l^2 m r^3\pm 40 l^2 r^4-20 r^6\right)\nonumber \right.\\&\left.+a_0^2 \left(20 \epsilon_0 \epsilon_1 l^4 m r^3\pm 10 \epsilon_0 \epsilon_1 l^2 r^6+4 \epsilon_0 l^4 m^2+4 \epsilon_0 l^4 m r-20 \epsilon_0 l^4 r^2\mp 50 \epsilon_0 l^2 m r^3\pm 50 \epsilon_0 l^2 r^4-26 \epsilon_0 r^6\right)+\mathcal{O}(a_0^3)\right]\,.
\end{align}
Integrating Eq.~\eqref{energyreps} by parts, we obtain 
\begin{align}
E&=\int_{r_1}^{r_2}dr  \left[\mathcal{H}''(r)-\mathcal{G}'(r)\right]\delta a(r)+\left[\left(\mathcal{G}(r)-\mathcal{H}'(r)\right)\delta a(r)+\mathcal{H}(r) \delta a'(r)\right]_{r_1}^{r_2}\,.
\end{align}
Using $\mathcal{H}''(r)=\mathcal{G}'(r)$, $\delta a'(r_2)=\delta a'(r_1)=0$, and
\begin{equation} \nonumber
	\mathcal{G}(r)-\mathcal{H}'(r)=-\epsilon_0 m-a_0 m \left(\epsilon_0^2+ \epsilon_1 \pm\frac{4}{l^2}\right)- 2 a_0^2 \epsilon_0 m\left(\epsilon_1\mp \frac{1}{l^2}\right)+\mathcal{O}(a_0^3)\,, 
\end{equation}
we get
\begin{equation}
	E=-\left[\epsilon_0 m+a_0 m \left(\epsilon_0^2+ \epsilon_1 \pm \frac{4}{l^2}\right)+ 2 a_0^2 \epsilon_0 m\left(\epsilon_1\mp \frac{1}{l^2}\right)\right]\Delta a+\mathcal{O}(a_0^3)\,.
\end{equation}

On the other hand, it is easily seen from \eqref{physical} that
\begin{equation}
\Delta M = \mp \frac{4 a_0 m}{l^2}\Delta a+ \mathcal{O}(a_0^3) \,.
\end{equation}
Finally, imposing $E=\Delta M$ as an equality of power series in $a_0$ we obtain $\epsilon_0=\epsilon_1=0$. 
%
%
%
\section{Test fields cannot destroy extremal Kerr-Newman-dS black holes}\label{section5}
In the previous sections we have shown that the timelike Killing vector field $K$ given by Eq.~\eqref{killing2} is the correct choice to compute the energy of a test field in a Kerr-Newman-de Sitter background, at least in what concerns its interaction with the black hole. On the other hand, it is well known that the null generator of the event horizon is $Z=K+\Omega Y$, where $\Omega$ is the thermodynamic angular velocity, that is, the angular velocity that occurs in the first law (see for instance \cite{DKKMT13, KS16}). Therefore, we can apply Theorem~4.1 in \cite{NQV16} to conclude that test fields cannot destroy extremal Kerr-Newman-dS black holes. More precisely, we have the following result.

\begin{Thm}
Test fields satisfying the null energy condition at the event horizon and appropriate boundary conditions at infinity cannot destroy extremal Kerr-Newman-dS black holes. More precisely, if an extremal black hole is characterized by the physical quantities $(M,J,Q)$, and absorbs energy, angular momentum and electric charge $(\Delta M,\Delta J,\Delta Q)$ by interacting with the test fields, then the metric corresponding to the physical quantities $(M + \Delta M, J + \Delta J, Q + \Delta Q)$ represents either a subextremal or an extremal black hole.
\end{Thm}
%
%
%
\section{Conclusions}\label{section6}
In this paper we have shown that the timelike Killing vector field $K$ given by Eq.~\eqref{killing2} gives the correct definition of energy for test fields propagating in the Kerr-Newman-dS spacetime. Additionally, we have confirmed that the timelike Killing vector field $K$ given by Eq.~\eqref{killing} gives the correct definition of energy for test fields propagating in the Kerr-Newman-AdS spacetime, as was already assumed in \cite{NQV16}. Moreover, using the general result in \cite{NQV16}, we proved that test fields cannot destroy extremal Kerr-Newman-dS black holes. 

The technique employed in this paper, namely allowing parameters in the metric to become functions in order to interpolate between black hole spacetimes with different physical masses, can be useful in other situations where the choice of the timelike Killing vector field with which to compute the energy of test fields is not clear. It is also possible that these ideas may play a role in determining an appropriate definition of mass for asymptotically de Sitter spacetimes.
%
%
%
\section*{Acknowledgments}
JN was partially funded by FCT/Portugal through UID/MAT/04459/2013 and grant (GPSEinstein) PTDC/MAT-ANA/1275/2014.
RV was supported by the FCT PhD scholarship SFRH/BD/128834/2017.
%
%
%


\begin{thebibliography}{BLCNR10}

\bibitem{W74}
R.\ Wald, \emph{Gedanken experiments to destroy a black hole}, Ann.\ of Phys.\ \textbf{83} (1974), 548--556.

\bibitem{P69}
R.\ Penrose, \emph{Gravitational collapse: the role of general relativity}, Riv.\ Nuovo Cim.\ \textbf{1} (1969), 252--276.

\bibitem{W97}
R.\ Wald, \emph{Gravitational collapse and cosmic censorship}, \href{https://arxiv.org/abs/gr-qc/9710068}{\texttt{arXiv:gr-qc/9710068}} (1997).

\bibitem{TdFC76}
K.\ Tod, F.\ de\ Felice and M.\ Calvani, \emph{Spinning test particles in the field of a black hole}, Nuovo Cim.\ B \textbf{34} (1976), 365--379.

\bibitem{Needham80}
T.\ Needham, \emph{Cosmic censorship and test particles}, Phys.\ Rev.\ D \textbf{22} (1980), 791--796.

\bibitem{Semiz11}
I.\ Semiz, \emph{Dyonic Kerr-Newman black holes, complex scalar field and cosmic censorship}, Gen.\ Rel.\ Grav.\ \textbf{43} (2011), 833--846.

\bibitem{Toth12}
G.\ Toth, \emph{Test of the weak cosmic censorship conjecture with a charged scalar field and dyonic Kerr–Newman black holes}, Gen.\ Rel.\ Grav.\ \textbf{44} (2015), 2019--2035.

\bibitem{DS13}
K.\ Duztas and I.\ Semiz, \emph{Cosmic censorship, black holes and integer-spin test fields}, Phys.\ Rev.\ D \textbf{88} (2013), 064043.

\bibitem{Duztas14}
K.\ Duztas, \emph{Electromagnetic field and cosmic censorship}, Gen.\ Rel.\ Grav.\ \textbf{46} (2014), 1709.

\bibitem{BCNR10}
M.\ Bouhmadi-Lopez, V.\ Cardoso, A.\ Nerozzi and J.\ Rocha, \emph{Black holes die hard: can one spin-up a black hole past extremality?}, Phys.\ Rev.\ D \textbf{81} (2010), 084051.

\bibitem{RV17}
K.\ Revelar and I.\ Vega, {\em Overcharging higher-dimensional black holes with point particles}, Phys.\ Rev.\ D {\bf 96} (2017) 064010.

\bibitem{ASZZ17}
J.\ An, J.\ Shan, H.\ Zhang and S.\ Zhao, {\em $5$-dimensional Myers-Perry Black Holes Cannot be Over-spun by Gedanken Experiments}, Phys.\ Rev.\ D {\bf 97} (2018) 104007.

\bibitem{GL16}
B.\ Gwak and B.\ Lee, {\em Cosmic censorship of rotating Anti-de Sitter black hole}, JCAP 02 (2016) 015.

\bibitem{RS14}
J.\ Rocha and R.\ Santarelli, \emph{Flowing along the edge: spinning up black holes in AdS spacetimes with test particles}, Phys.\ Rev.\ D \textbf{89} (2014), 064065.

\bibitem{Gwak18}
B.\ Gwak, {\em Weak cosmic censorship conjecture in Kerr-(anti-)de Sitter black hole with scalar field}, JHEP 1809 (2018) 081. 

\bibitem{Hubeny99}
V.\ Hubeny, \emph{Overcharging a black hole and cosmic censorship}, Phys.\ Rev.\ D \textbf{59} (1999), 064013.

\bibitem{MS07}
G.\ Matsas and A.\ Silva, \emph{Overspinning a nearly extreme charged black hole via a quantum tunneling process}, Phys.\ Rev.\ Lett.\ \textbf{99} (2007), 181301.

\bibitem{JS09}
T.\ Jacobson and T.\ Sotiriou, \emph{Over-spinning a black hole with a test body}, Phys.\ Rev.\ Lett.\ \textbf{103} (2009), 141101.

\bibitem{SS11}
A.\ Saa and R.\ Santarelli, \emph{Destroying a near-extremal Kerr-Newman black hole}, Phys.\ Rev.\ D \textbf{84} (2011), 027501.

\bibitem{Hod08}
S.\ Hod, \emph{Weak cosmic censorship: as strong as ever}, Phys.\ Rev.\ Lett.\ \textbf{100} (2008), 121101.

\bibitem{BCK10}
E.\ Barausse, V.\ Cardoso and G.\ Khanna, {\em Test bodies and naked singularities: is the self-Force the cosmic censor?}, Phys.\ Rev.\ Lett.\ {\bf 105} (2010) 261102.

\bibitem{ZVPH13}
P.\ Zimmerman, I.\ Vega, E.\ Poisson and R.\ Haas, {\em Self-force as a cosmic censor}, Phys.\ Rev.\ D {\bf 87} (2013) 041501(R).

\bibitem{SPAJ15}
S.\ Shaymatov, M.\ Patil, B.\ Ahmedov and P.\ Joshi, {\em Destroying a near-extremal Kerr black hole with a charged particle: can a test magnetic field serve as a cosmic censor?}, Phys. Rev. D {\bf 91} (2015) 064025.

\bibitem{CBSM15}
M.\ Colleoni, L.\ Barack, A.\ Shah and M.\ van de Meent, {\em Self-force as a cosmic censor in the Kerr overspinning problem}, Phys.\ Rev.\ D {\bf 92} (2015) 084044.

\bibitem{NQV16}
J.\ Nat\'{a}rio, L.\ Queimada and R.\ Vicente, \emph{Test fields cannot destroy extremal black holes}, Class.\ Quant.\ Grav.\ \textbf{33} (2016), 175002.

\bibitem{SW17}
J.\ Sorce and R.\ Wald, {\em Gedanken Experiments to Destroy a Black Hole II: Kerr-Newman Black Holes Cannot be Over-Charged or Over-Spun}, Phys.\ Rev.\ D {\bf 96} (2017) 104014.

\bibitem{Wang01}
X.\ Wang, {\em The Mass of Asymptotically Hyperbolic Manifolds}, J.\ Differ.\ Geom.\ {\bf 57} (2001) 273-299.

\bibitem{CN02}
P.\ Chrusciel and G.\ Nagy, {\em The mass of spacelike hypersurfaces in asymptotically anti-de Sitter space-times}, Adv.\ Theor.\ Math.\ Phys.\ {\bf 5} (2002) 697-754.

\bibitem{Olea05}
R.\ Olea, \emph{Mass, angular momentum and thermodynamics in four-dimensional Kerr-AdS black holes}, J.\ High Energy Phys., JHEP06(2005).

\bibitem{MO15}
B.\ McInnes and Y.\ Ong, {\em A Note on Physical Mass and the Thermodynamics of AdS-Kerr Black Holes}, JCAP 11 (2015) 004.

\bibitem{KT02}
D.\ Kastor and J.\ Traschen, {\em A positive energy theorem for asymptotically de Sitter spacetimes}, Class.\ Quant.\ Grav.\ {\bf 19} (2002) 5901-5920.

\bibitem{LXZ10}
M.\ Luo, N.\ Xie and X.\ Zhang, \emph{Positive mass theorems for asymptotically de Sitter spacetimes}, Nucl.\ Phys.\ B \textbf{825} (2010) 98-118.

\bibitem{MTW73}
C.\ Misner, K.\ Thorne and J.\ A. Wheeler, \emph{Gravitation}, Freeman, 1973.

\bibitem{W84}
R.\ Wald, \emph{General relativity}, University of Chicago Press, 1984.

\bibitem{CCK00}
M.\ Caldarelli, G.\ Cognola and D.\ Klemm, \emph{Thermodynamics of Kerr-Newman-AdS black holes and conformal field theories}, Class.\ Quant.\ Grav.\ \textbf{17} (2000), 399--420.

\bibitem{Gwak18b}
B.\ Gwak, {\em Thermodynamics and Cosmic Censorship Conjecture in Kerr-Newman-de Sitter Black Hole}, Entropy {\bf 20} (2018) 855.

\bibitem{DKKMT13}
B.~Dolan, D.~Kastor, D.~Kubiznak, R.~Mann and J.~Traschen, {\em Thermodynamic Volumes and Isoperimetric Inequalities for de Sitter Black Holes}, Phys.\ Rev.\ D {\bf 87} (2013) 104017.

\bibitem{KS16}
D.~Kubiznak and F.~Simovic, {\em Thermodynamics of horizons: de Sitter black holes and reentrant phase transitions}, Class.\ Quant.\ Grav.\ {\bf 33} (2016) 245001.

\end{thebibliography}
\end{document}